\documentclass[preprint,english,preprintnumbers,amsmath,amssymb,floatfix]{revtex4}
\usepackage[T1]{fontenc}
\usepackage[latin9]{inputenc}
\usepackage{color}
\usepackage{array}
\usepackage{amstext}
\usepackage{graphicx}
\usepackage{esint}

\makeatletter

\@ifundefined{textcolor}{}
{%
 \definecolor{BLACK}{gray}{0}
 \definecolor{WHITE}{gray}{1}
 \definecolor{RED}{rgb}{1,0,0}
 \definecolor{GREEN}{rgb}{0,1,0}
 \definecolor{BLUE}{rgb}{0,0,1}
 \definecolor{CYAN}{cmyk}{1,0,0,0}
 \definecolor{MAGENTA}{cmyk}{0,1,0,0}
 \definecolor{YELLOW}{cmyk}{0,0,1,0}
 }

\@ifundefined{definecolor}
 {\usepackage{color}}{}
\@ifundefined{definecolor}
 {\usepackage{color}}{}
\makeatother

\makeatother

\usepackage{babel}

\begin{document}
% Some useful tex commands
%
%\newcommand{\qsq}{\ensuremath{Q^2} }
%\newcommand{\gevsq}{\ensuremath{\mathrm{GeV}^2} }
%\newcommand{\et}{\ensuremath{E_t^*} }
%\newcommand{\rap}{\ensuremath{\eta^*} }
%\newcommand{\gp}{\ensuremath{\gamma^*}p }
%\newcommand{\dsiget}{\ensuremath{{\rm d}\sigma_{ep}/{\rm d}E_t^*} }
%\newcommand{\dsigrap}{\ensuremath{{\rm d}\sigma_{ep}/{\rm d}\eta^*} }
\newcommand{\MSbar}{\ensuremath{\overline{\text{MS}}\ }}

\title{NNLO compatibility between pQCD theory and phenomenology in determination of the $b$-quark pole and \MSbar running masses}

\author{A.~Vafaee}
\email[]{vafaee.phy@gmail.com}
\affiliation{Department of Physics, Ferdowsi University of Mashhad, P.O.Box 1436, Mashhad, Iran}
\affiliation{Iran's National Elites Foundation, P. O. Box 14578-93111, Tehran, Iran}

\author{K.~Javidan}
\email[]{javidan@ferdowsi.um.ac.ir}
\affiliation{Department of Physics, Ferdowsi University of Mashhad, P.O.Box 1436, Mashhad, Iran}

\author{A.~B.~Shokouhi}
\email[]{shokouhi.phy@gmail.com}
\affiliation{Independent researcher, P. O. Box 11155-811, Tehran, Iran}

\date{\today}

\begin{abstract}
This contribution attempts to determine the $b$-quark pole mass $M_b$ and \MSbar running mass $\overline{m}_b$ with two different approaches at the next-to-next-to-leading order (NNLO) corrections. At the first approach, we derive a relation between the $b$-quark pole mass $M_b$ and its \MSbar running mass $\overline{m}_b$ at the NNLO corrections based on the perturbative Quantum Chromo Dynamics (pQCD) predictions.  At the second approach, we extract numerical values of the $b$-quark pole and \MSbar running masses based on the NNLO phenomenology of H1 and ZEUS Collaborations combined beauty vertex production experimental data. Then we discuss about the compatibility between the pQCD theory results and phenomenology approach in determination of the $b$-quark pole and \MSbar running masses at the NNLO corrections. Also, we investigate the role and influence of the $b$-quark mass as an extra degree of freedom added to the input parameters of the Standard Model Lagrangian, on the improvement of the uncertainty band of the proton parton distribution functions (PDFs) and particularly on the gluon distribution.  
\end{abstract}

%\pacs{12.38.Aw}
%\keywords{}

\maketitle

\section{\label{introduction}Introduction}
Measurements of the open $b$-quark production in deep inelastic scattering (DIS) of ${e^\pm}p$ at HERA provide an important test of pQCD within the framework of Standard Model and are used to constrain the proton PDFs.
 
 Phenomenologically, the proton parton distribution functions as a non-calculable part of the factorization theorem are classically extracted by QCD fitting of a parameterized standard functional form with experimental data from deep inelastic $e^{\pm}p$ scattering at HERA collider~\cite{Abramowicz:2015mha,H1:2018flt}. The experimental data extracted from DIS of $e^{\pm}p$ play central role in probing of the internal structure of the proton. The inclusive neutral current (NC) and charged current (CC) cross sections at HERA contain contributions from all active quark and antiquark flavors and it should be noted that up to $45$~\% of these contributions are originated from events with charm and beauty quarks in the final states~\cite{Gluck:1993dpa,Thorne:1997ga}.

 Measurements of open $b$-quark production in DIS of ${e^\pm}p$ at HERA provide an important test of pQCD theory within the Standard Model and is used to constrain the proton PDFs. On the other hand, the $b$-quark mass as an extra degree of freedom added to the input parameters of the Standard Model Lagrangian play central role in high energy physics phenomenology~\cite{Gribov:1972rt}. The $b$-quark mass as an important pQCD parameter is significantly larger than the QCD scale parameter $\Lambda_{QCD} \sim 250$~MeV and accordingly it is known as a heavy-quark which is now kinematically accessible at HERA. At the leading order (LO) or $0$-loop corrections, the dominant processes for the $b$-quark production at HERA are generally known as the boson-gluon fusion (BGF) reactions $\gamma g\rightarrow b\overline{b}$ which are strongly sensitive to the gluon content of the proton~\cite{Aktas:2006vs}.  

Within the pQCD framework, the ratio of photon couplings corresponding to a heavy-quark is given by $f(h) \sim \frac{Q_h^2}{\Sigma{Q_q^2}}$ where $Q_h^2$ denotes the electric charge squared of a heavy-quark and $Q_q^2$ ($q=u,d,s,c,b$) refers to the electric charge squared of the kinematically accessible quark flavors at HERA. Accordingly for the $b$-quark we may write: $f(b) \sim \frac{Q_b^2}{\Sigma{Q_q^2}} = \frac{1}{9} / \frac{11}{9} = \frac{1}{11}$ or $f(b) \sim 0.09$, which clearly shows that up to $9$~\% of the HERA inclusive cross sections are originated from processes with the $b$-quark in the final states. Therefore investigation of the role and influence of the $b$-quark experimental data on the proton PDFs and determination of the $b$-quark pole mass $M_b$ and \MSbar running mass $\overline{m}_b$ as an extra degree of freedom in the input parameters of the Standard Model Lagrangian play central role in many of pQCD analysis~\cite{Abramowicz:2011kj}.

 It should be noted that in the pQCD framework the $b$-quark mass has been generated via a spontaneous
symmetry breaking mechanism which in turn caused by a non-zero vacuum expectation value of a Higgs boson field. However, the $b$-quark mass remains as an extra free parameter of the the Standard Model Lagrangian and should be determined by comparing of phenomenology of experimental data with theoretical predictions of the pQCD theory. These are our main motivations to develop this article. This contribution attempts to determine the $b$-quark pole mass $M_b$ and \MSbar running mass $\overline{m}_b$, using phenomenology of experimental data and pQCD theory predictions at the next-to-next-to-leading order. Then we discuss for the first time about the compatibility between pQCD theory results and phenomenology approach in determination of the $b$-quark pole and \MSbar running masses at the NNLO corrections. Also we investigate the role and influence of the $b$-quark mass as an extra pQCD parameter on the improvement of the uncertainty band of the proton PDFs and particularly on the gluon distribution.

 This paper is organized as follows. In Sec.~\ref{dis}, we describe the neutral and charged currents of deep inelastic ${e^\pm}p$ scattering cross sections and introduce the inclusive cross section of the $b$-quark production, as well. Heavy-flavor treatments to insert of heavy-quark contributions in deep inelastic electro-production proton structure functions are explained in In Sec.~\ref{heavy}. We explain and discuss about the $b$-quark pole and \MSbar running masses based on the perturbative quantum chromodynamics theory predictions in Sec.~\ref{Theory}. In Sec.~\ref{methodology}, the systematic uncertainties and our QCD analysis set-up are discussed. We present our results in Sec.~\ref{results} and finally, we conclude with a summary and conclusion in Sec.~\ref{Summary}.

\section{\label{dis}The inclusive cross section of b-quark production}

The HERA particle accelerator at the Deutsches Elektronen-Synchrotron (DESY) as a large QCD laboratory, study both neutral and charged currents of $e^{\pm}p$ collisions and its data cover a wide range of phase space in Bjorken $x$ scale and negative four-momentum squared of the virtual photon $Q^2$~\cite{Abramowicz:2015mha}.

The NC interactions cross sections have been published for
$4.5 \cdot 10^{-4} \leq Q^2 \leq 5.0 \cdot 10^{4} $\,GeV$^2$
and  $6.0 \cdot 10^{-7} \leq x \leq 6.5 \cdot 10^{-3}$ 
at values of the inelasticity $5.0 \cdot 10^{-3} \leq y = \dfrac{Q^2}{sx} \leq 9.5 \cdot 10^{-1}$~. The reduced NC unpolarized deep inelastic ${e^\pm}p$ scattering cross sections at the centre-of-mass energies up to $\sqrt{s} \simeq 320\,$GeV after correction for Quantum Electro Dynamics (QED) radiative effects can be expressed in terms of the NC generalized structure functions $\tilde{F_2}$, $x\tilde{F_3}$ and $\tilde{F_{\rm L}}$ as follows~\cite{Abramowicz:2015mha}:
\begin{eqnarray}
\label{eq:NC}
   \sigma_{r,NC}^{{\pm}}&=&   \frac{d^2\sigma_{NC}^{e^{\pm} p}}{d{x}dQ^2} \frac{Q^4 x}{2\pi \alpha^2 (1 + (1-y)^2)}\\
   &=& \tilde{F_2} \mp \frac{(1 - (1-y)^2)}{(1 + (1-y)^2)} x\tilde{F_3} -\frac{y^2}{(1 + (1-y)^2)} \tilde{F_{\rm L}}~. \nonumber
\end{eqnarray}

Similarly the CC interactions cross sections have been
published for $2.0 \cdot 10^{2} \leq Q^2 \leq 5.0 \cdot 10^{4} $\,GeV$^2$
and  $1.3 \cdot 10^{-2} \leq x \leq 4.0 \cdot 10^{-1}$ 
at values of the inelasticity $3.7 \cdot 10^{-4} \leq y = \dfrac{Q^2}{sx} \leq 7.6 \cdot 10^{-3}$~. The reduced cross sections for inclusive unpolarized CC ${e^\pm}p$ scattering are defined in terms of CC structure functions $W_2^{\pm}$, $W_3^{\pm}$ and $W_L^{\pm}$ as follows~\cite{Abramowicz:2015mha}:  
\begin{eqnarray}
 \label{eq:CC}
\sigma_{r,CC}^{\pm} &=&\frac{2\pi x}{G^2_F} \left[\frac{M^2_W+Q^2}{M^2_W}\right]^2 \frac{d^2\sigma_{CC}^{e^{\pm} p}}{d{x}dQ^2}\\
     &=& \frac{(1 + (1-y)^2)}{2}  W_2^{\pm} \mp \frac{(1 - (1-y)^2)}{2}x  W_3^{\pm} - \frac{y^2}{2} W_L^{\pm}~~.  \nonumber
\end{eqnarray}
  
In Eqs.~(\ref{eq:NC}) and ~(\ref{eq:CC}), the quantity $\alpha$ refers to the fine-structure constant which is defined at zero momentum transfer frame and $G_F$ refers to the Fermi constant which is related to the weak coupling constant $g$ and electromagnetic coupling constant $e$ by $G^2_F = \frac{e^2}{4\sqrt{2}{\sin ^2\theta_W}M^2_W} = \frac{g^2}{4M_W}$~\cite{Abramowicz:2015mha}. More details about proton NC and CC generalized structure functions can be found in the Ref.~\citep{Vafaee:2017nze}.

 The NC measurements of deep inelastic $e^{\pm}p$ scattering at HERA for beauty contribution to the inclusive proton cross sections $F_2^{b\overline{b}}$ have been studied by H1 and ZEUS Collaborations in the range of virtuality of the exchanged photon $2.5$ GeV$^2 \le Q^2 \le 2000$ GeV$^2$ and Bjorken $x$ scale of $3.0 \cdot 10^{-5} \le x \le 5.0 \cdot 10^{-2}$~\cite{H1:2018flt}.
% based on a dataset with an integrated luminosity of ${\cal L}=189$~${\rm pb}^{-1}$~\cite{h}.

 In analogy to the inclusive NC and CC deep inelastic ${e^\pm}p$ scattering cross sections, the reduced cross sections for the $b$-quark production in deep inelastic $e^{\pm}p$ scattering measurements can be expressed in terms of the $b$-quark contributions to the inclusive structure functions $F_2^{b\bar b}$, $x  F_3^{b\bar b}$ and $F_L^{b\bar b}$ as follows: 
\begin{eqnarray}
 \label{eq:NCheavy}
	\sigma_{red}^{b\bar{b}} &=& 
	          \frac{d\sigma^{b\bar{b}}(e^{\pm} p)}{d{x}dQ^2} \frac{Q^4 x}{2 \pi \alpha^2 (1 + (1-y)^2)}\\
	       &=& F_2^{b\bar b} \mp \frac{(1 - (1-y)^2)}{(1 + (1-y)^2)}x  F_3^{b\bar b} - \frac{y^2}{(1 + (1-y)^2)}  F_L^{b\bar b}~.  \nonumber
\end{eqnarray}
 
 Within the quark parton model (QPM) framework where $Q^2\ll M_Z^2$ the parity-violating structure function $xF_3$ can be neglected and accordingly the reduced cross sections for the $b$-quark contributions can be expressed by 
\begin{eqnarray}
 \label{eq:RNCheavy}
	\sigma_{red}^{b\bar{b}} &=& 
	          \frac{d\sigma^{b\bar{b}}(e^{\pm} p)}{d{x}dQ^2} \frac{Q^4 x}{2 \pi \alpha^2 (1 + (1-y)^2)} \\
	      &=& F_2^{b\bar b} - \frac{y^2}{(1 + (1-y)^2)}  F_L^{b\bar b}~.   \nonumber
\end{eqnarray}
More details can be found in the Ref.~\cite{H1:2018flt}.

The double-differential cross sections $\frac{d^2\sigma^{b\bar{b}}(e^{\pm} p)}{d{x}dQ^2}$ for the production of a $b$-quark may be written in terms of the $b$-quark contributions to the proton structure functions $F^{\rm b\overline{b}}_2(x,Q^2)$ and 
$F^{\rm b\overline{b}}_{\rm L}(x,Q^2)$ as follows:
 \begin{equation}
\frac{d^2\sigma^{b\bar{b}}(e^{\pm} p)}{d{x}dQ^2} = \frac{2\pi \alpha^2(Q^2)}{xQ^4} ( [1+(1-y)^2]F^{\rm b\overline{b}}_2 -y^2F_{\rm L}^{\rm b\overline{b}} )\ ,
\label{eqn:dcs}
\end{equation}
where  $y$ is the lepton inelasticity. It should be noted that the superscripts $\rm b\overline{b}$ indicate the presence of a $b$-quark pair in the final state. The double-differential cross sections $\frac{d^2\sigma^{b\bar{b}}(e^{\pm} p)}{d{x}dQ^2}$ is given at the Born level without electroweak radiative and QED corrections, except for the running electromagnetic coupling $\alpha(Q^2)$. More details can be found in the Ref.~\cite{H1:2018flt}.

In this pQCD analysis, we perform three different fits entitled: PPDFs, HBPoleMass and HBRunMass so that in throughout of this article the words PPDFs, HBPoleMass and HBRunMass  refer as follows:
\begin{itemize}
\item {PPDFs:} The PPDFs analysis indicates the determination of our central proton PDFs (PPDFs) based on the seven set of HERA run I and II combined data~\cite{Abramowicz:2015mha}.

\item {HBPoleMass:} The HBPoleMass analysis indicates the determination of the $b$-quark pole mass $M_b$ using HERA I and II combined~\cite{Abramowicz:2015mha} and H1 and ZEUS Collaboration beauty combined production~\cite{H1:2018flt} data sets.

\item {HBRunMass:} The HBRunMass analysis indicates the determination of the $b$-quark \MSbar running mass $\overline{m}_b$ using HERA I and II combined~\cite{Abramowicz:2015mha} and H1 and ZEUS Collaboration beauty combined production~\cite{H1:2018flt} data sets.

\end{itemize}

\section{\label{heavy} Heavy-flavor treatments}
To insert of the heavy-quark contributions in deep inelastic electro-production proton structure functions and separating the proton structure functions into the  proton PDFs and calculable processes based on the QCD factorization theorem, there exist various heavy-flavor treatments. Within the pQCD framework there are generally two different classes of heavy-flavor treatments for deep inelastic ${e^\pm}p$ scattering cross sections, which are known as the so-called the fixed flavor number (FFN) and variable flavor number (VFN) schemes~\cite{Martin:2006qz,Aivazis:1993kh}.

 The FONLL method was originally introduced to describe the transverse-momentum distribution of heavy-quarks in hadronic collisions and is developed based on the following formula:

\begin{equation}
  \label{eq:merge}
  \mbox{FONLL}=\mbox{FO}\;+\left( \mbox{RS}\; -\; \mbox{FOM0}\right)\;
\times G(m,p_T)\;,
\end{equation}
where FO and RS stand for fixed-order and resummed approaches, respectively, FOM0 stands for a massless limit of the fixed-order calculation (an approximation of FO where only logarithmic
mass terms are retained) and $G(m,p_T)$ is an arbitrary function with only this restriction that: $G(m,p_T) \rightarrow 1$ as $m/p_T\,\to\,0$, up to terms suppressed by powers of $m/p_T$.

 The FONLL approach contains of various variants and this NNLO QCD analysis, based on our QCD set-up and methodology which will be described in detail in Sec.~\ref{methodology}, we use FONLL-C and FONLL-C RUNM ON as two different variants of FONLL approach which have been provided by APFEL C$++$ code only at NNLO corrections~\cite{Bertone:2013vaa} to inclusion of the $b$-quark contributions to the proton structure function at the NNLO QCD correction order. A very good study with more details about inclusion of heavy-quark mass contributions to deep-inelastic proton structure functions based on the FONLL method can be found in the Ref.~\cite{Forte:2010ta}.

\section{\label{Theory} Perturbative Quantum Chromodynamics theory predictions}
One of the main features of QCD theory is color confinement postulate (the QCD short range feature) which says all hadron states and physical observables such as currents, energies, momenta and masses are color-singlet. It should be noted that the above postulate is just a kincmatical constraint to eliminate colored states. There is, however a hope that the quark confinement may be the natural dynamical consequence of quantum chromodynamics theory.

 Because of color confinement feature of QCD, free quarks are unobservable and accordingly there are different definitions for the $b$-quark mass such as pole mass and \MSbar running mass.

 Physically, the definition of the $b$-quark mass comes from its contribution as an extra free parameter in QCD Lagrangian as an one degree of freedom of non-Abelian gauge field theory and its exact value depends on the specific renormalization scheme.

 In the on-shell renormalization scheme, the $b$-quark mass is defined as the pole of the $b$-quark propagator and known as the $b$-quark pole mass $M_b$. The $b$-quark pole mass definition is same as the typical definition of the lepton mass. 

 In the \MSbar scheme, the $b$-quark mass is defined as a scale-dependent perturbative running parameter and is called the $b$-quark \MSbar running mass. The $b$-quark \MSbar running mass definition is the same as the definition of the running strong coupling $\alpha_s(Q^2)$.
  
%  Each of these definitions for the $b$-quark mass has their own advantages and disadvantages. The $b$-quark pole mass is a gauge invariant quantity and is well defined in each finite order of pQCD theory, but this definition  as the pole of the $b$-quark propagator involves some contribution from the non-perturbative region and accordingly suffers from an intrinsic uncertainty of order $\frac{\Lambda_{QCD}}{m_b}$, where as we previously mentioned the $\Lambda_{QCD} \sim 250$~MeV refers to the QCD scale parameter. On the other hand, since the \MSbar running mass is defined at the renormalization scale $\mu_r$ where $\mu_r \gg \Lambda_{QCD} $, therefor the $b$-quark \MSbar running mass definition avoids the intrinsic uncertainty of pole mass definition.
  
Within the pQCD framework the connection between renormalized and unrenormalized (bare) quark mass is given by

\begin{eqnarray}
\label{eq:con1}
   m_{0} && = Z^{\overline{\rm MS}}_m \overline{m}_b,  \\
   m_{0} && = Z^{\rm OS}_m M_b,
\label{eq:con2}
\end{eqnarray}
where $m_0$ is the unrenormalized or bare $b$-quark mass and 
  $Z^{\overline{\rm MS}}_m$ and $Z^{\rm OS}_m$ are  renormalization factors for the $b$-quark mass in the 
   $\overline{\rm MS}$ and the on-shell schemes, respectively.
   
From Eqs.~(\ref{eq:con1}),~(\ref{eq:con2}) we may write the relation between the $b$-quark pole mass $M_b$ and \MSbar running mass $\overline{m}_b$ as follows:
\begin{eqnarray}
\label{eq:ratio}
   \frac{\overline{m}_b}{M_b} = \frac{Z^{\rm OS}_m}{Z^{\overline{\rm MS}}_m}.
\end{eqnarray}

Now, it is clear from Eq.~(\ref{eq:ratio}) to extract relation between the $b$-quark pole mass $M_b$ and its \MSbar running mass $\overline{m}_b$ at the NNLO, it is enough to determine $Z^{\overline{\rm MS}}_m$ and $Z^{\rm OS}_m$ renormalization factors in both the \MSbar and the on-shell schemes at the NNLO.

Within the pQCD framework the mass renormalization factor $Z^{\overline{\rm MS}}_m$ in the $\overline{\rm MS}$-scheme is given by:

\begin{eqnarray}
\label{eq:zms}
  Z^{\overline{\rm MS}}_m = 1 
+\sum \limits_{i=1}^{\infty} C_i \left ( \frac {\alpha_s(\mu)}{\pi} \right )^i,
\end{eqnarray}
with
\begin{eqnarray}
\label{eq:cparms}
C_1 && = -\frac {1}{\varepsilon}, \nonumber \\
C_2 && = \frac {1}{\varepsilon^2}
         \left ( \frac {15}{8} - \frac {1}{12} N_f \right )
+ \frac {1}{\varepsilon}\left (  - \frac {101}{48} + \frac {5}{72}N_f \right ),
\nonumber \\
\end{eqnarray}
where $N_f$ is the number of different fermion flavors and $\varepsilon$ is the dimensional regularization parameter which is related to the the space-time dimension $D$ by  $\varepsilon = \frac{4-D}{2}$.

To determine the renormalization factor $Z^{\rm OS}_m$ in the on-shell scheme we start from the perturbative quark propagator which is defined as follows:
\begin{eqnarray}
\label{eq:propagator}
\hat S_F(p) = \frac {i}{\hat p - m_0 + \hat \Sigma(p,M_b)},
\end{eqnarray}
where $\hat \Sigma(p,M_b)$ is the one particle irreducible $b$-quark self-energy which is parameterized as follows:
\begin{eqnarray}
\label{eq:irreducible}
\hat \Sigma(p,M_b) = M \Sigma_1(p^2,M_b) + (\hat p - M_b)\Sigma_2(p^2,M_b).
\end{eqnarray}
 Since the $b$-quark pole mass $M_b$ corresponds to the position of the pole of the $b$-quark propagator we may write
\begin{eqnarray}
\label{eq:ZOSsigma}
    Z^{\rm OS}_m = 1 + \left. \Sigma_1(p^2,M_b)\right|_{p^2=M_b^2}~,
\end{eqnarray}
which is the simplest formula for the renormalization factor $Z^{\rm OS}_m$ in the on-shell scheme.

Now, having computed the NNLO contribution to $Z^{\rm OS}_m$ and using  Eqs.~(\ref{eq:ratio}),~(\ref{eq:zms}), 
we can obtain a NNLO relation between the $b$-quark pole mass $M_b$ and \MSbar running mass $\overline{m}_b$ in terms of the color factors as follows:
\begin{eqnarray}
\label{eq:relationbpar}
{\overline m_b}(M_b) && = M_b \left[ 
1 - C_F \left ( \frac {\alpha_s}{\pi} \right )
+ C_F \left ( \frac {\alpha_s}{\pi} \right )^2 
\left (C_F\, d^{(2)}_1  
\right. \right. \nonumber \\
&& \left. \left.
 + C_A\, d^{(2)}_2 
+ T_RN_L\,d^{(2)}_3
+ T_RN_H \, d^{(2)}_4
\right ) \right],
\end{eqnarray}
where:
\begin{itemize}
\item $C_F$ is the Casimir operator of the fundamental representation of the color gauge SU($3$) group.
\item $C_A$ is the Casimir operator of the adjoint representation of the color gauge SU($3$) group.
\item $T_R$ denotes the trace normalization of the 
fundamental representation.
\item $N_L$ refers to the number of massless quark flavors.
\item $N_H$ refers to the number of quark flavors with a pole mass equal to $M_b$.
\item  $\alpha_s \equiv \alpha_s^{(N_L+N_H)}(M_b)$ refers to the ${\overline {\rm MS}}$ 
strong coupling which is renormalized at the scale of the pole mass $\mu = M_b$ in the pQCD theory with $N_L+N_H$ active flavors.
\end{itemize}

From Eqs.~(\ref{eq:relationbpar}), we may obtain the following results for the coefficients 
 $d^{(n)}_k$:
 
\begin{eqnarray}
\label{eq:mainresult}
d^{(2)}_1 && =  \frac {7}{128} - \frac {3}{4}\zeta_3 
 + \frac {1}{2}\pi^2 \log2 - \frac {5}{16}\pi^2, 
\nonumber \\
d^{(2)}_2 && =   
 - \frac {1111}{384} + \frac {3}{8}\zeta_3 - \frac {1}{4}\pi^2 \log2 
 + \frac {1}{12} \pi^2, 
\nonumber \\
d^{(2)}_3 && =  \frac {71}{96} + \frac {1}{12}\pi^2, 
\nonumber \\
 d^{(2)}_4 && =  \frac {143}{96} - \frac {1}{6}\pi^2. 
\nonumber 
\end{eqnarray}

Now, if we insert the standard values of the pQCD color factors:  $C_F=4/3,~C_A=3,~~T_R=1/2$ and setting the number of heavy-flavors to $N_H = 1$, one may finds the following result:

\begin{eqnarray}
\label{eq:main}
&& {\overline m_b}(M_b) = M_b \left[  
 1 - \frac {4}{3} \left (\frac {\alpha_s}{\pi} \right )
        + \left (\frac {\alpha_s}{\pi} \right )^2 \left (
       N_L \left ( \frac {71}{144} + \frac {\pi^2}{18} \right )
\right. \right. \nonumber \\
&& \left. \left. - \frac {3019}{288} + \frac {1}{6}\zeta_3 
         - \frac {\pi^2}{9}\log 2 - \frac {\pi^2}{3}
 \right ) \right], 
\end{eqnarray}
or numerically we may find:
\begin{eqnarray}
\label{eq:msthrpole}
&& {\overline m_b}(M_b) = M_b \left[  
1 - \frac{4}{3} \left ( \frac {\alpha_s}{\pi} \right )
+\left ( \frac {\alpha_s}{\pi} \right )^2
\left ( 1.0414~N_L  -14.3323 \right )
 \right],
\end{eqnarray}
or:
\begin{eqnarray}
\label{eq:resultmm}
&& M_b = \overline {m}_b(\overline {m}_b) \left[  
1 + \frac{4}{3} \left ( \frac {\bar \alpha_s}{\pi} \right )
+\left ( \frac {\bar \alpha_s}{\pi} \right )^2
\left ( - 1.0414~N_L  + 13.4434 \right )
 \right],
\end{eqnarray}
with $\bar \alpha_s \equiv \alpha_s(\overline {m}_b)$.

It should be noted that at the leading order (LO) or $0$-loop calculation, $ \left ( \frac {\alpha_s}{\pi} \right ) \rightarrow 0$ and accordingly the difference between the $b$-quark pole mass and its \MSbar running mass vanishes.

\section{\label{methodology} QCD analysis Set-UP and Systematic uncertainties}
In this contribution, we made three different fits to determine the $b$-quark pole mass $M_b$ and \MSbar running mass $\overline{m}_b$ based on the following QCD set-up: 
\begin{itemize}
\item To choose the PDFStyle and parametrize the PDFs, we use generic HERAPDF functional form:
\begin{equation}
 xf(x) = A x^{B} (1-x)^{C} (1 + D x + E x^2)~~,
\label{eqn:pdf}
\end{equation}
with $14$ free central fit parameters and $1$ extra free parameter $m_b$ at the starting scale of the QCD evolution $Q^2_0= 1.9$ GeV$^2$~\cite{Vafaee:2017nze}.

\item In order to fit of experimental data on theory, we us the xFitter package as a powerful QCD open source framework~\cite{xFitter,Vafaee:2019nmo,Vafaee:2019yec,Shokouhi:2018gie,Vafaee:2018ehy,Vafaee:2018abd,Vafaee:2017jnt,Vafaee:2016jxl}.

\item Evolution of the parametrized PDFs has been done based on the DGLAP collinear evolution with QCDNUM~\cite{Botje:2010ay} and APFEL packages~\cite{Bertone:2013vaa} corresponding to FONLL-C and FONLL-C RUNM ON schemes, respectively.

\item We choose the lower band of the $b$-quark mass as $M_b = 4.192$~GeV~and then varied in steps of 0.01.

\item To include light flavor contribution, we set the renormalization and factorization scales as $\mu_r = \mu_f = Q$, while for the $b$-quark contribution we use the typical definition as: $\mu_r = \mu_f = \mu_r = \sqrt{Q^2 + 4m_b^2}$.

\item The strangeness suppression factor and the strong coupling constant fixed to $f_{s}=0.4$ and $\alpha_s^{{\rm NNLO}}(M^2_Z) = 0.118$~, respectively.
   
\end{itemize}

\section{\label{results}Results}
In the Table~(\ref{tab:data}), we show data sets used in our QCD analysis, correlated ${\chi^2}$ and extracted ${\frac{\chi^2_{Total}}{dof}}$ corresponding to HBPoleMass and HBRunMass analysis, respectively.

 According to the numerical results from Table~(\ref{tab:data}), the best quality of the fit in determination of the $b$-quark pole mass $M_b$ and \MSbar running mass $\overline{m}_b$ is $\frac{\chi^2_{Run}}{dof}={\frac{1364}{1157}}$ corresponding to HBRunMass analysis. Also, according to relative improvement of $\chi^2-$function which is defined by $\frac{\chi^2_{Pole}-\chi^2_{Run}}{\chi^2_{Pole}}$, we get an improvement up to $\frac{1.187-1.178}{1.187}$ $\sim 0.8$~\%  in the quality of the fit for determination of the $b$-quark \MSbar running mass $\overline{m}_b$ relative to the $b$-quark pole mass $M_b$.
 
\begin{table}[h]
\begin{center}
\begin{tabular}{|l|c|c|c|c|}
\hline
\hline
\multicolumn{3}{|c|}{HERA run I $+$ II combined and H1 and ZEUS Collaboration beauty combined data}    \\ \hline
 {Experiment} & {$~~~~${HBPoleMass}$~~~~$} & {HBRunMass} \\ \hline
  HERA I$+$II CC $e^{+}p$~\cite{Abramowicz:2015mha} & 51 / 39& 50 / 39 \\ 
  HERA I$+$II CC $e^{-}p$~\cite{Abramowicz:2015mha} & 49 / 42& 49 / 42 \\ 
  HERA I$+$II NC $e^{-}p$~\cite{Abramowicz:2015mha} & 218 / 159& 217 / 159 \\ 
  HERA I$+$II NC $e^{+}p$ 460~\cite{Abramowicz:2015mha} & 215 / 204& 215 / 204 \\ 
  HERA I$+$II NC $e^{+}p$ 575~\cite{Abramowicz:2015mha} & 212 / 254& 211 / 254 \\
  HERA I$+$II NC $e^{+}p$ 820~\cite{Abramowicz:2015mha} & 62 / 70& 62 / 70 \\ 
  HERA I$+$II NC $e^{+}p$ 920~\cite{Abramowicz:2015mha} & 419 / 377& 413 / 377 \\ \hline
 {H1 and ZEUS beauty combined}~\cite{H1:2018flt} & 15 / 27& 15 / 27 \\ \hline
 {Correlated ${\chi^2}$} & 133& 132 \\\hline
{${\frac{\chi^2_{Total}}{dof}}$}  & $\frac{\chi^2_{Pole}}{dof}={\frac{1374}{1157}}$ &  $\frac{\chi^2_{Run}}{dof}={\frac{1364}{1157}}$  \\ \hline
\hline
    \end{tabular}
\vspace{-0.0cm}
\caption{\label{tab:data}{Data sets, correlated ${\chi^2}$ and extracted ${\frac{\chi^2_{Total}}{dof}}$ corresponding to HBPoleMass and HBRunMass analysis, respectively.}}
\vspace{-0.4cm}
\end{center}
\end{table}

 In the Table~(\ref{tab:par}), we present NNLO numerical values of $15$ fit parameters and their uncertainties, including $14$ free central PDF parameters and $1$ extra $m_b$ parameter corresponding to HBPoleMass and HBRunMass analysis.

 According to the numerical results from Table~(\ref{tab:par}) for the $b$-quark pole mass $M_b = 4.66 \pm 0.14$~GeV and \MSbar running mass $ \overline{m}_b = 4.40 \pm 0.10$~GeV, we obtain up to $\sim 4.0$~\% pure improvement in the uncertainty value of the $b$-quark \MSbar running mass $\overline{m}_b$ relative to the $b$-quark pole mass $M_b$.  
 
\begin{table}[h]
\begin{center}
\begin{tabular}{|l|c|c|c|c|}
\hline
\hline
\multicolumn{3}{|c|}{{Numerical values of fit parameters corresponding to HBPoleMass and HBRunMass analysis}}    \\ \hline
 {Parameter} & {$~~~~${HBPoleMass}$~~~~$} & {HBRunMass} \\ \hline
  ${B_{u_v}}$ & $0.846 \pm 0.039$& $0.848 \pm 0.039$ \\ 
  ${C_{u_v}}$ & $4.465 \pm 0.076$& $4.468 \pm 0.077$ \\ 
  $E_{u_v}$ & $11.4 \pm 1.6$& $11.4 \pm 1.6$ \\ \hline
  ${B_{d_v}}$ & $1.04 \pm 0.10$& $1.04 \pm 0.11$ \\ 
  $C_{d_v}$ & $4.24 \pm 0.39$& $4.21 \pm 0.40$ \\ \hline
  $C_{\bar{U}}$ & $7.25 \pm 0.97$& $7.22 \pm 0.98$ \\ 
  $D_{\bar{U}}$ & $8.5 \pm 2.2$& $7.7 \pm 2.1$ \\ 
  $A_{\bar{D}}$ & $0.1651 \pm 0.0092$& $0.1806 \pm 0.0095$ \\  
  $B_{\bar{D}}$ & $-0.1810 \pm 0.0068$& $-0.1695 \pm 0.0065$ \\ 
  $C_{\bar{D}}$ & $5.28 \pm 0.96$& $5.43 \pm 0.98$ \\ \hline
  $B_g$ & $0.12 \pm 0.12$& $0.12 \pm 0.14$ \\ 
  $C_g$ & $5.36 \pm 0.82$& $5.59 \pm 0.90$ \\ 
  $A_g'$ & $2.26 \pm 0.40$& $2.30 \pm 0.44$  \\ 
  ${B_g'}$ & $0.005 \pm 0.058$& $0.020 \pm 0.069$ \\ \hline
  $m_b$ & pole mass $M_b = 4.66 \pm 0.14$ & \MSbar running mass $ \overline{m}_b = 4.40 \pm 0.10$  \\  \hline 
\hline
    \end{tabular}
\vspace{-0.0cm}
\caption{\label{tab:par}{The NNLO numerical values of $15$ fit parameters and their uncertainties, including $14$ free central PDF parameters and $1$ extra $m_b$ parameter corresponding to HBPoleMass and HBRunMass analysis.}}
\vspace{-0.4cm}
\end{center}
\end{table}

 In Sec.~\ref{Theory}, we extracted the relation between the $b$-quark pole mass $M_b$ and \MSbar running mass $\overline{m}_b$ at the NNLO of pQCD framework as follows:
 
\begin{eqnarray}
\label{eq:msthrpole}
&& {\overline m_b}(M_b) = M_b \left[  
1 - \frac{4}{3} \left ( \frac {\alpha_s}{\pi} \right )
+\left ( \frac {\alpha_s}{\pi} \right )^2
\left ( 1.0414~N_L  -14.3323 \right )
 \right]. \nonumber
\end{eqnarray}

 Now, if we insert our phenomenology results for the $b$-quark pole and \MSbar running masses from Table~(\ref{tab:par}) into the Eq.~\ref{eq:msthrpole} (which comes from pQCD theory predictions) we get:

\begin{eqnarray}
\label{eq:t}
4.40 &\sim& 4.66 \left[  
1 - \frac{4}{3} \left ( \frac {0.118}{3.141} \right )
+\left ( \frac {0.118}{3.141} \right )^2
\left ( 1.0414\times 3  -14.3323 \right )
 \right]~, \\
% 4.40 &\sim& 4.66 \times 0.94~, \\
4.40 &\sim& 4.38~, \nonumber
\end{eqnarray}
where according to our methodology in Sec.~\ref{methodology}, we set the strong coupling constant at $M^2_Z$ scale to $\alpha_s^{{\rm NNLO}}(M^2_Z) = 0.118$.
  
 If we qualify the error as: $ \bigtriangleup x = \left|x_f - x_i\right|$, we see that the difference of our numerical results extracted based on the phenomenology of experimental data for the $b$-quark pole mass $M_b$ and \MSbar running mass $\overline{m}_b$ with the pQCD theory prediction at the NNLO corrections are less than $ \left|4.40 - 4.38\right| = 0.02$. In other words, the compatibility of our numerical results with the pQCD theory predictions at the NNLO corrections and with precision of $1$ part in $10^2$ are up to approximately $99.98$~\%  which show an excellent agreement between our phenomenology analysis results with pQCD theory. Furthermore, the comparison of our numerical results with the measurements from the PDG~\cite{Agashe:2014kda} world average, shows a very good agreement with the expected $b$-quark pole masses.

\begin{figure*}
\includegraphics[width=0.49\textwidth]{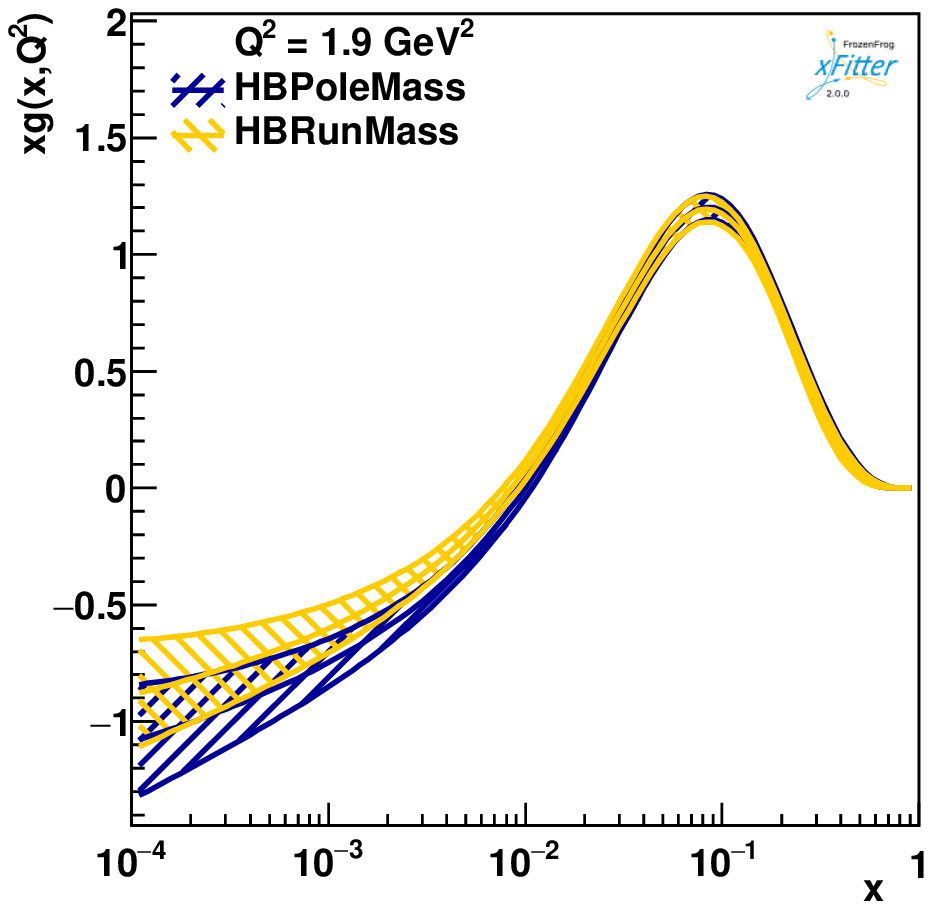}
\includegraphics[width=0.49\textwidth]{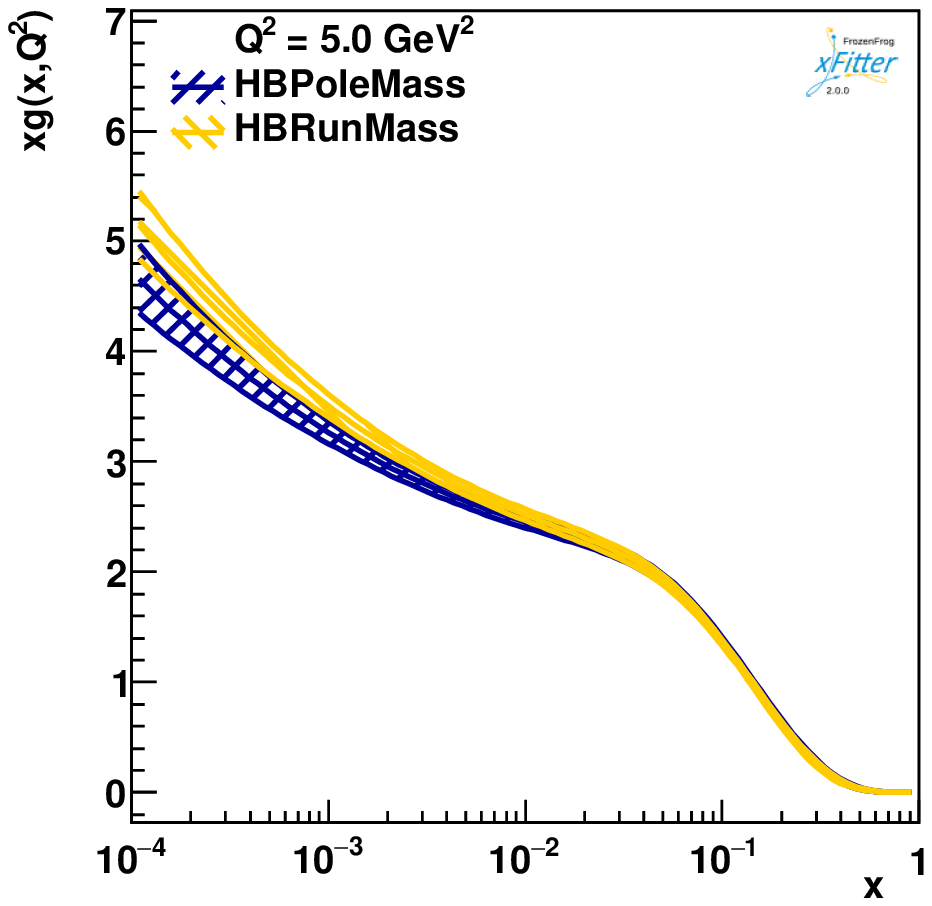}

\includegraphics[width=0.49\textwidth]{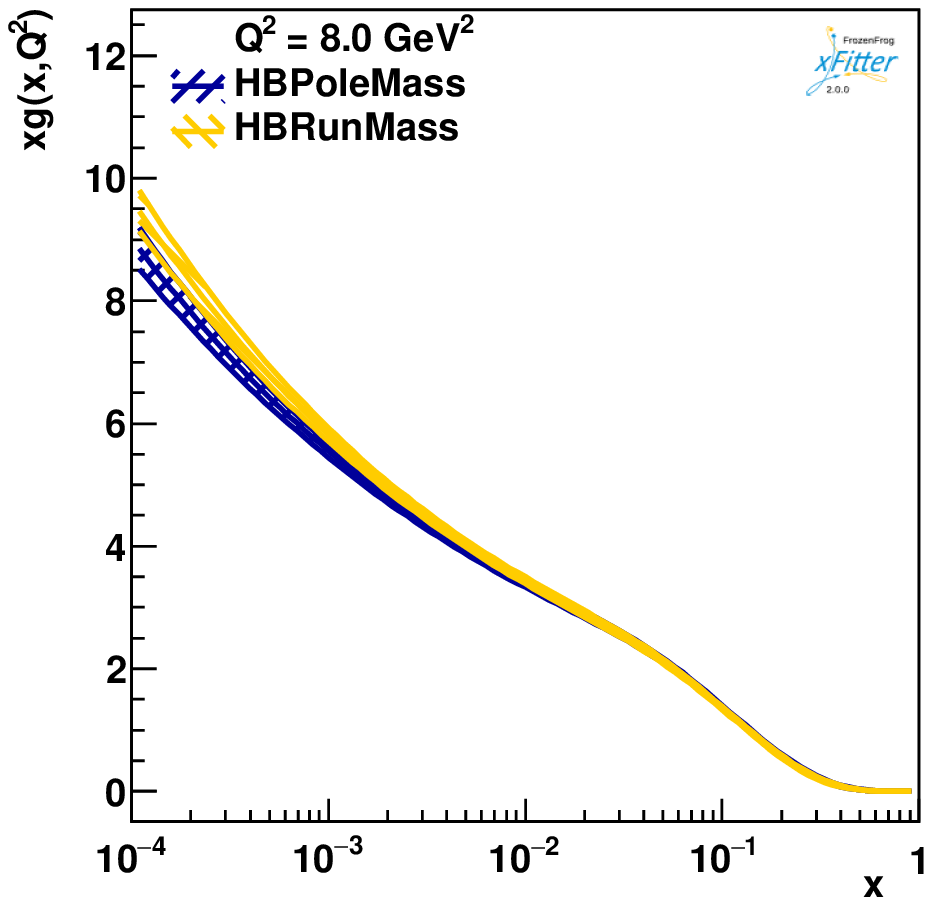}
\includegraphics[width=0.49\textwidth]{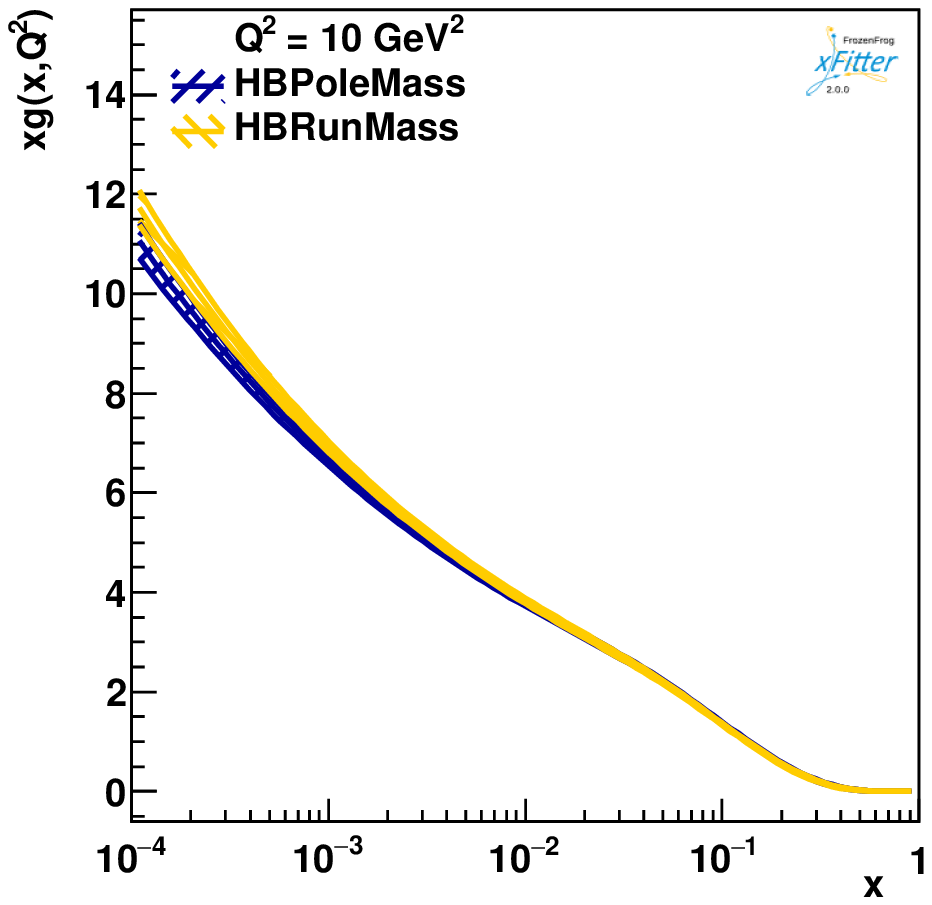}
\caption{Comparison of pure impact of the $b$-quark pole mass $M_b$ (blue color) and \MSbar running mass $\overline{m}_b$ (yellow color) on the gluon distribution as a function of $x$ at $Q^2 = 1.9, 5.0, 8.0$ and $10$~GeV$^2$.}
\label{fig:1}
\end{figure*}

  In the Fig.~(\ref{fig:1}), we compare the pure impact of the $b$-quark pole mass $M_b$ and \MSbar running mass $\overline{m}_b$ on the gluon distribution as a function of $x$ at $Q^2 = 1.9, 5.0, 8.0$ and $10$~GeV$^2$.

  As we expected from numerical results of Table~(\ref{tab:par}), the gluon distribution is sensitive to the $b$-quark mass, when it is considered as an extra free parameter in pQCD framework.

\begin{figure*}
\includegraphics[width=0.49\textwidth]{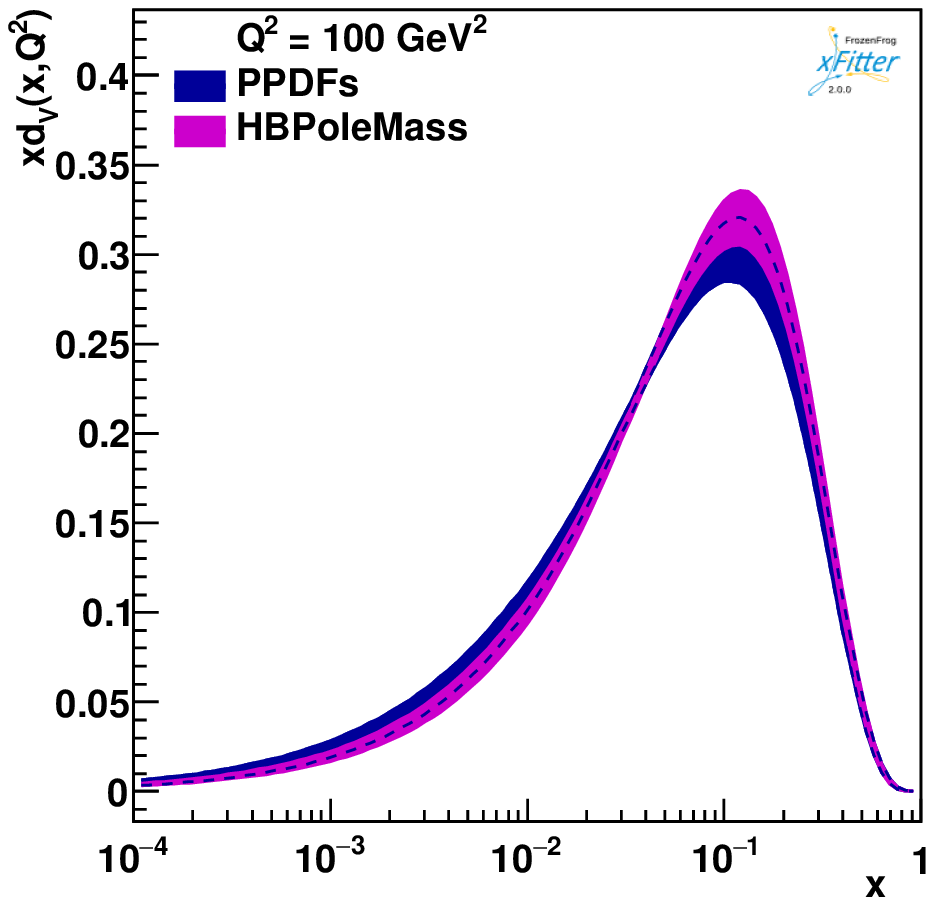}
\includegraphics[width=0.49\textwidth]{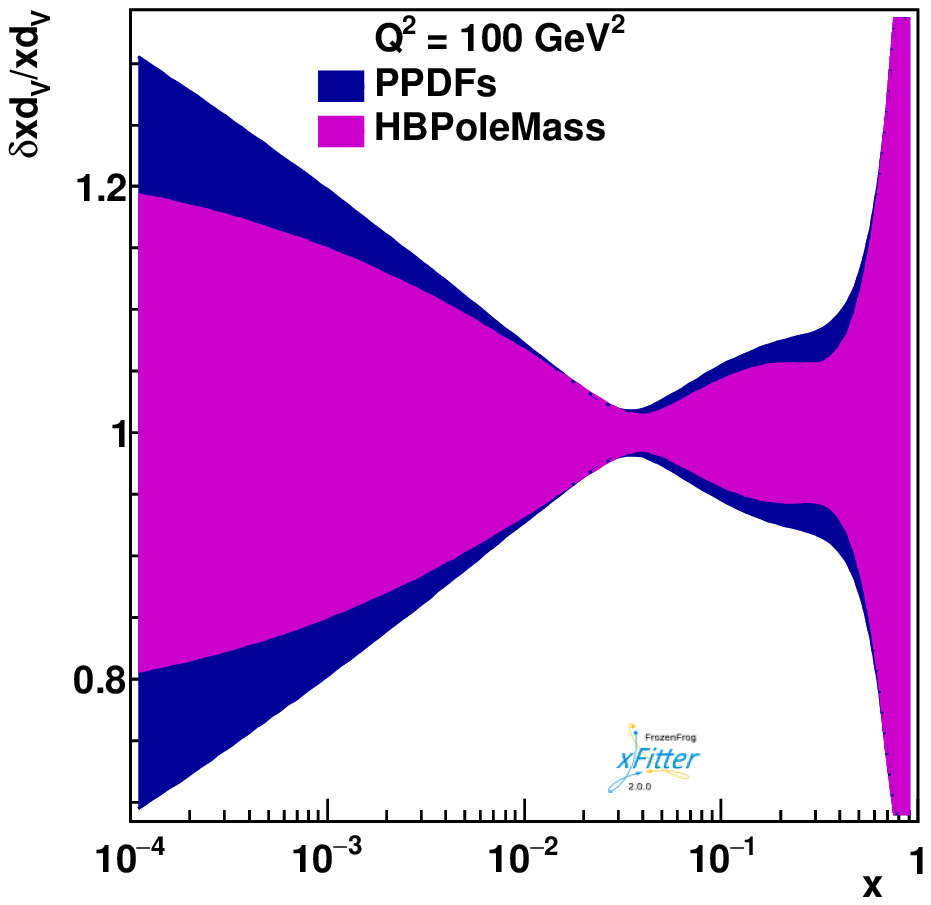}
\caption{Comparison of the $d$-valence and $d_v$-ratio of the proton PDFs without (PPDs analysis with blue color) and with (HBPoleMass analysis with purple color) inclusion of the $b$-quark pole mass $M_b$ as an extra degree of freedom added to the input parameters of the Standard Model Lagrangian.}
\label{fig:2}
\end{figure*}

The $d$-valence and $d_v$-ratio of the proton PDFs without (PPDs analysis with blue color) and with (HBPoleMass analysis with purple color) inclusion of the $b$-quark pole mass $M_b$ as an extra degree of freedom added to the input parameters of the Standard Model Lagrangian is shown in the Fig.~(\ref{fig:2}).

\begin{figure*}
\includegraphics[width=0.49\textwidth]{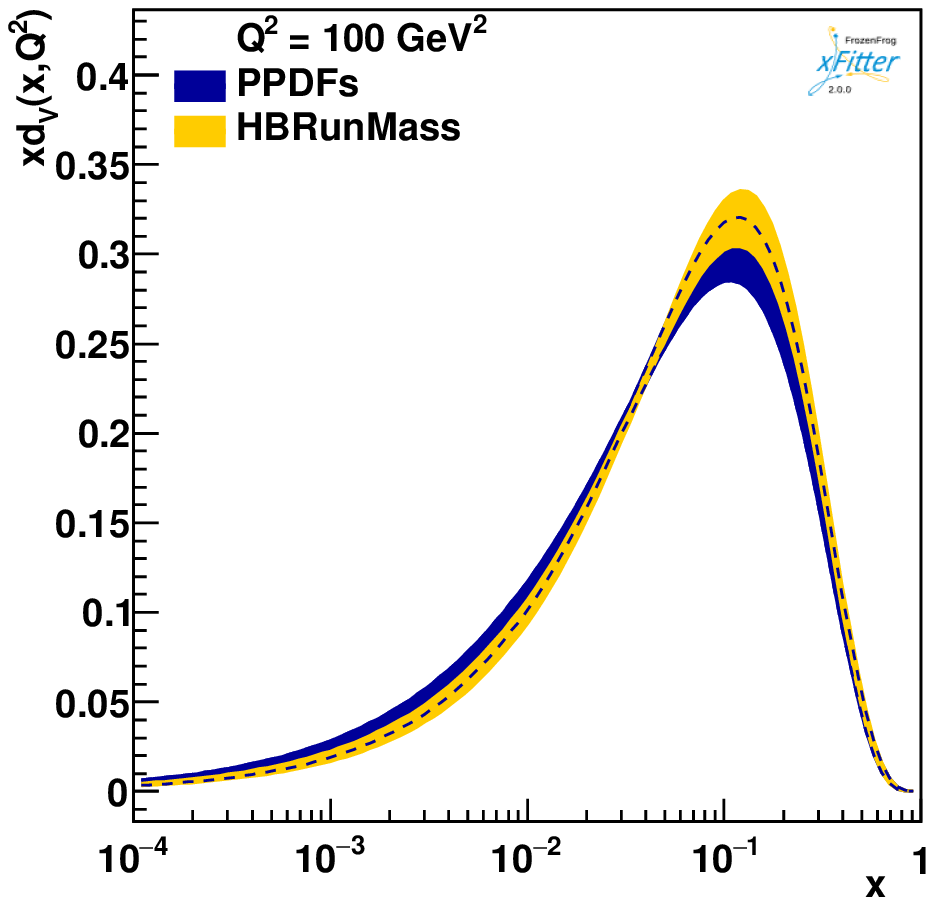}
\includegraphics[width=0.49\textwidth]{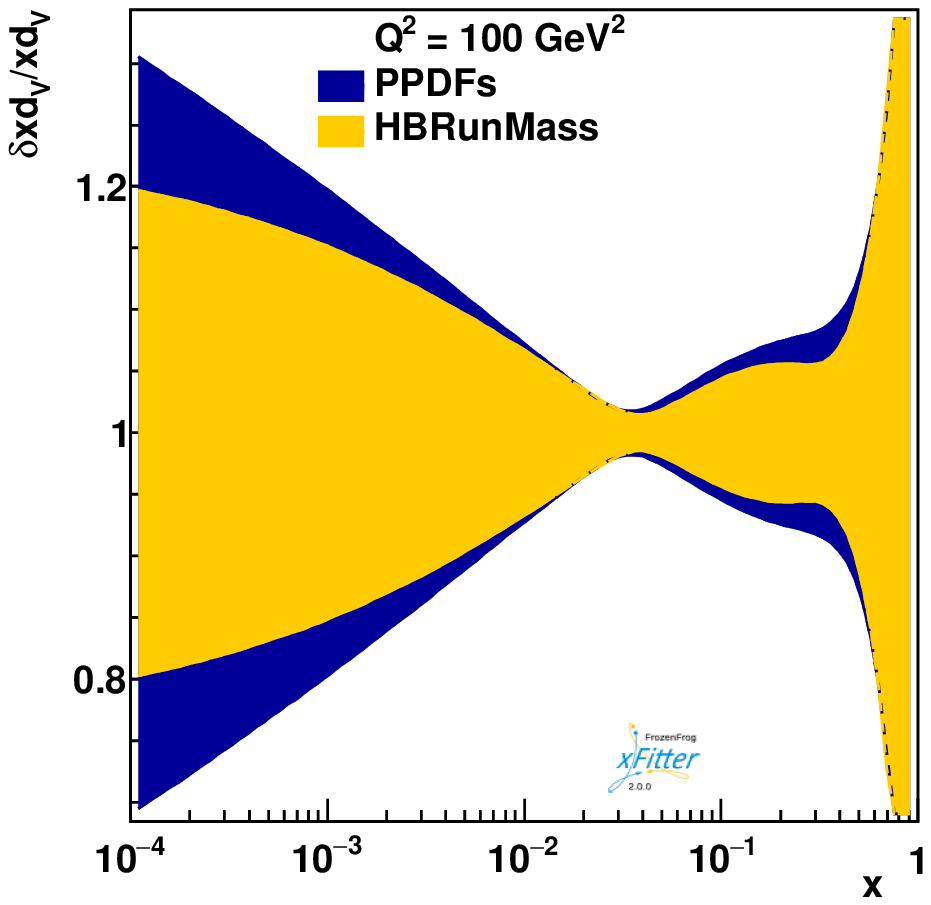}
\caption{Comparison of the $d$-valence and $d_v$-ratio of the proton PDFs without (PPDs analysis with blue color) and with (HBRunMass analysis with yellow color) inclusion of the $b$-quark \MSbar running mass $\overline{m}_b$ as an extra degree of freedom added to the input parameters of the Standard Model Lagrangian.}
\label{fig:3}
\end{figure*}

In the Fig.~(\ref{fig:3}), we compare the $d$-valence and $d_v$-ratio of the proton PDFs without (PPDs analysis with blue color) and with (HBRunMass analysis with yellow color) inclusion of the $b$-quark \MSbar running mass $\overline{m}_b$ as an extra degree of freedom added to the input parameters of the Standard Model Lagrangian.

From Figs.~(\ref{fig:2}) and (\ref{fig:3}), we see the dramatic impact of the $b$-quark pole and \MSbar running masses at the NNLO corrections on both $d$-valence and $d$-valence ratio of the proton central PDFs, which comes from strong correlation between proton PDFs and  the $b$-quark mass when it is considered as an extra degree of freedom added to the input parameters of the pQCD Lagrangian.
  
\begin{figure*}
\includegraphics[width=0.49\textwidth]{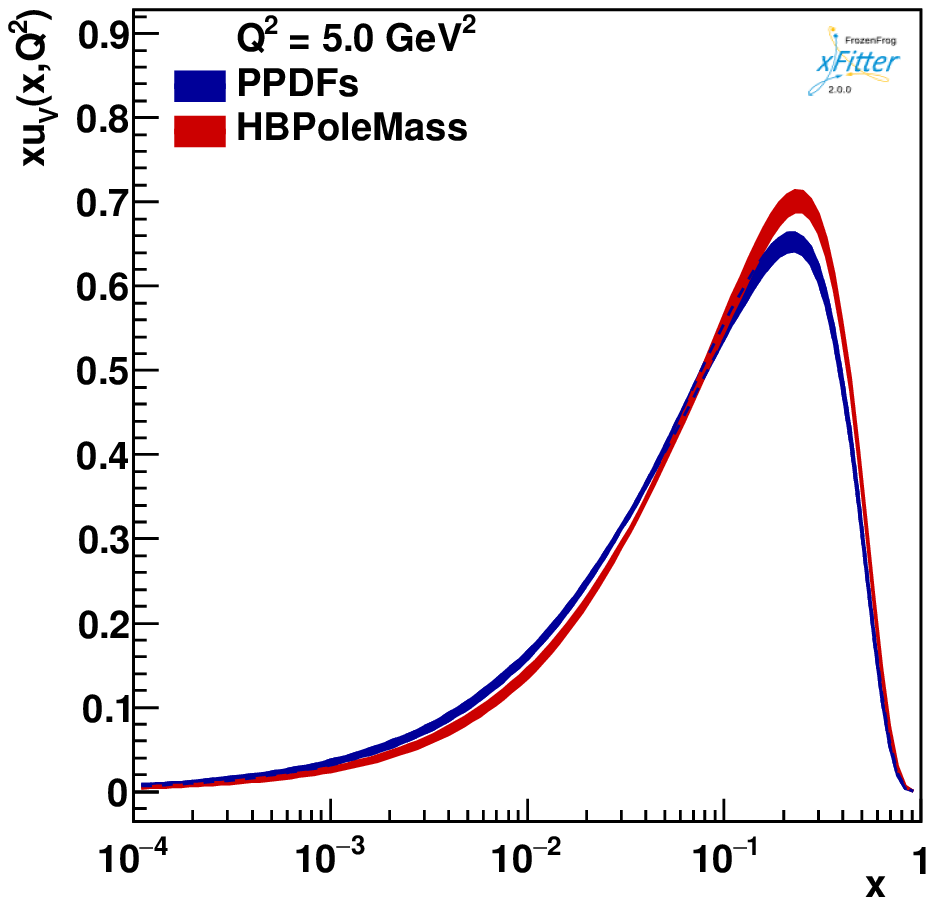}
\includegraphics[width=0.49\textwidth]{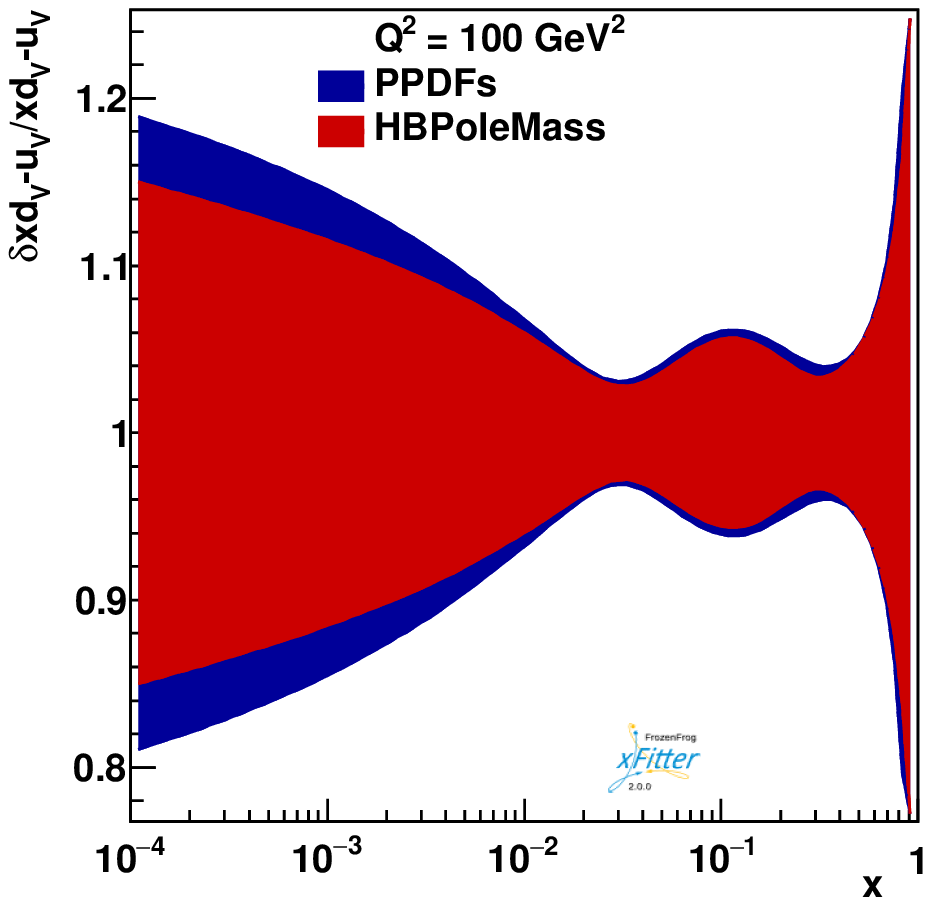}
\caption{Comparison of the $u$-valence and $u_v-d_v$-ratio of the proton PDFs without (PPDs analysis with blue color) and with (HBPoleMass analysis with red color) inclusion of the $b$-quark pole mass $M_b$ as an extra degree of freedom added to the pQCD Lagrangian.}
\label{fig:4}
\end{figure*}

The $u$-valence and $u_v-d_v$-ratio of the proton PDFs without (PPDs analysis with blue color) and with (HBPoleMass analysis with red color) inclusion of the $b$-quark pole mass $M_b$ as an extra degree of freedom added to the pQCD Lagrangian is shown in the Fig.~(\ref{fig:4}).
 
\begin{figure*}
\includegraphics[width=0.49\textwidth]{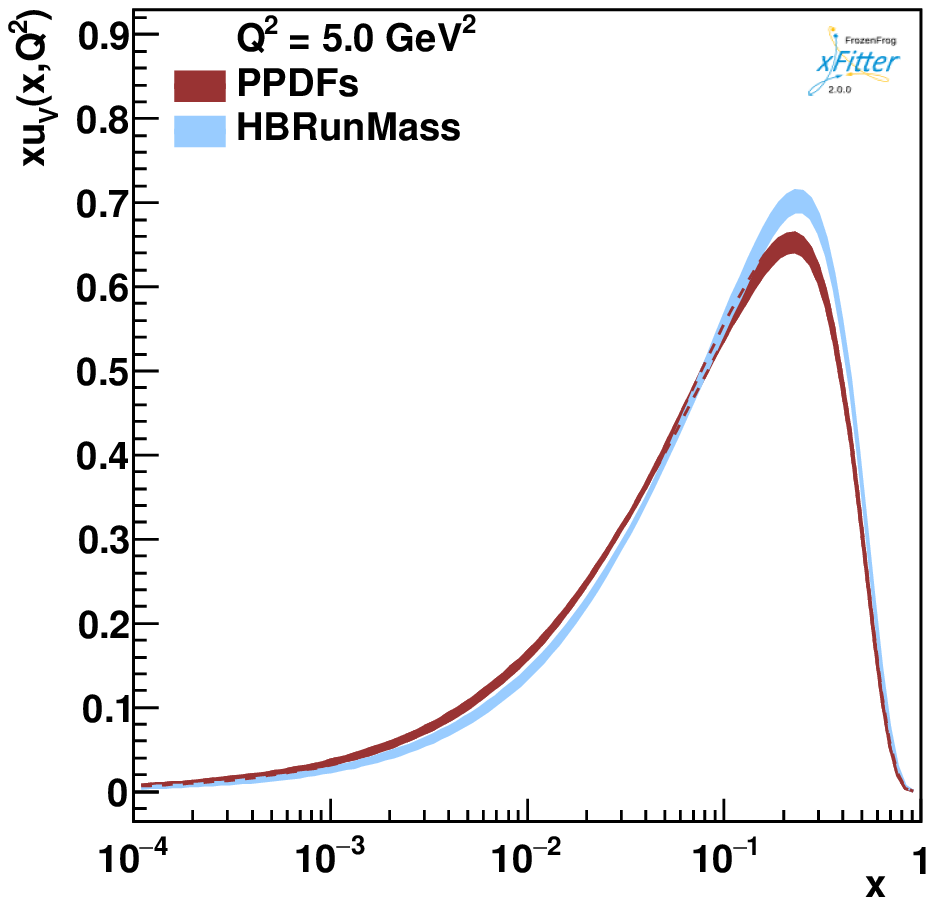}
\includegraphics[width=0.49\textwidth]{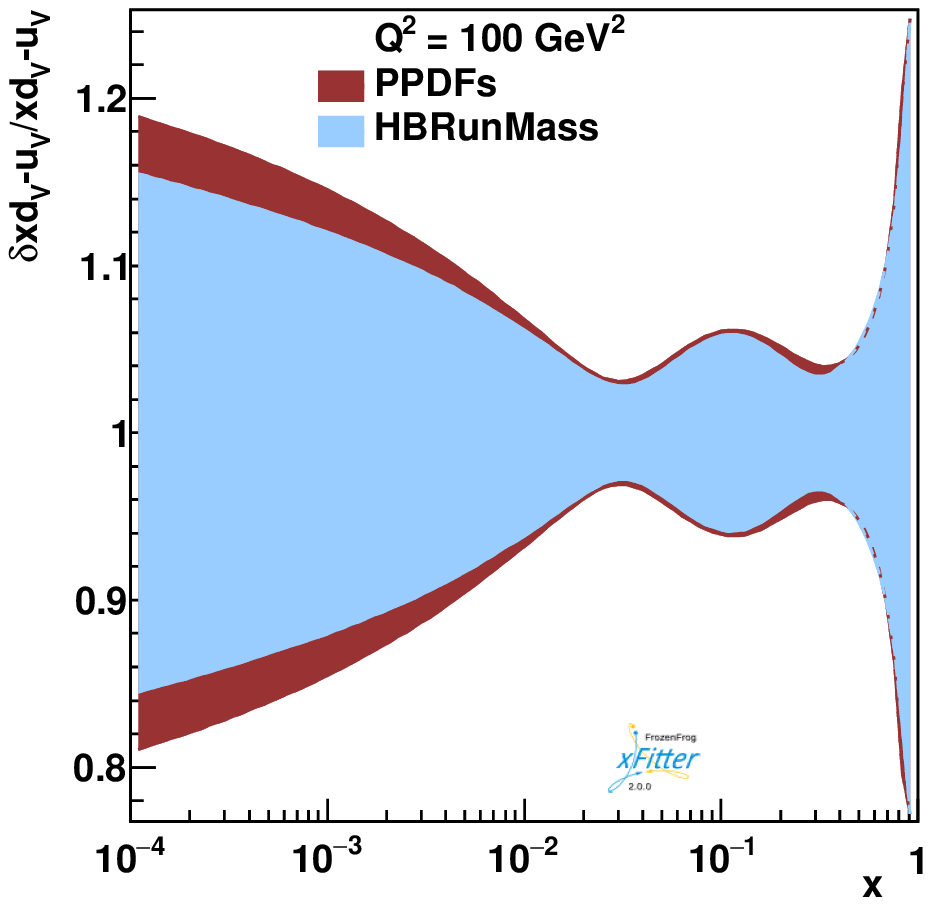}
\caption{Comparison of the $u$-valence and $u_v-d_v$-ratio of the proton PDFs without (PPDs analysis with red color) and with (HBRunMass analysis with blue color) inclusion of the $b$-quark \MSbar running mass $\overline{m}_b$ as an extra degree of freedom added to the pQCD Lagrangian.}
\label{fig:5}
\end{figure*}

In the Fig.~(\ref{fig:5}), we compare the $u$-valence and $u_v-d_v$-ratio of the proton PDFs without (PPDs analysis with red color) and with (HBRunMass analysis with blue color) inclusion of the $b$-quark \MSbar running mass $\overline{m}_b$ as an extra degree of freedom added to the pQCD Lagrangian.

\clearpage
\section{\label{Summary}Summary and Conclusion}

\begin{itemize}
\item We determine the $b$-quark pole mass $M_b$ and \MSbar running mass $\overline{m}_b$ with two different approaches at NNLO corrections. At the first approach, we derive a relation between the $b$-quark pole mass $M_b$ and its \MSbar running mass $\overline{m}_b$ based on the pQCD predictions.  At the second approach we extract numerical values of the $b$-quark pole and \MSbar running masses based on the phenomenology of experimental data.

\item Comparison of the numerical results for the $b$-quark pole mass $M_b$ and its \MSbar running mass $\overline{m}_b$ at the NNLO corrections extracted from phenomenological base approach with pQCD theory prediction shows an excellent compatibility between these two different approaches.

\item Based on the latest measurements of open $b$-quark production in deep inelastic scattering of ${e^\pm}p$ at HERA, which have been reported for the first time by H1 and ZEUS Collaborations combined beauty vertex production data, we show an excellent compatibility up to approximately $99.98$~\% between the pQCD theory predictions and the phenomenology approach results in determination of the $b$-quark pole and \MSbar running masses at the NNLO corrections.

\item Heavy-quark production measurements may be used to constrain important QCD parameters, such as the $b$-quark pole and \MSbar running masses. Also such measurements has some important consequences for the determination of other  pQCD parameters like the strong coupling constant $\alpha_s(M^2_Z)$. This NNLO QCD analysis reveals the role and influence of the $b$-quark pole and \MSbar running masses as an extra degree of freedom added to the input parameters of the Standard Model Lagrangian in the improvement of the uncertainty band of the proton PDFs and particularly for gluon distribution and some of its ratios.

\end{itemize}

\clearpage
\section{Acknowledgments}
We gratefully acknowledge V. Radescu for guidance and useful discussions about PDFs and xFitter. We are grateful to Prof. M. Botje from Nikhef, Science Park, Amsterdam, the Netherlands for providing the QCDNUM package as a very fast QCD evolution program. We are also grateful to Prof. F. Olness from SMU for developing invaluable heavy-flavor schemes as implemented in the xFitter package. We would like to thank Dr. Francesco Giuli, Dr. Ivan Novikov, Dr. Oleksandr Zenaiev and Dr. Sasha Glazov from xFitter developer group for guidance and  technical support. We would like to appreciate Mrs. Malihe Shokouhi for spending time and careful reading the draft version of this manuscript. This work is related to the ``Special Support Program for the Promotion of Scientific Authority'' in Ferdowsi University of Mashhad.

\end{document}